# Dependence of electronic structure of SrRuO$_3$ and the degree of correlation on cation off-stoichiometry

Wolter Siemons[1,2,*], Gertjan Koster[1,*], Arturas Vailionis[1], Hideki Yamamoto[1,3], Dave H.A. Blank[2] and Malcolm R. Beasley[1]

[1] *Geballe Laboratory for Advanced Materials, Stanford University, Stanford, California, 94305, United States of America*

[2] *Faculty of Science and Technology and MESA+ Institute for Nanotechnology, University of Twente, P.O. Box 217, 7500 AE, Enschede, The Netherlands*

[3] *NTT Basic Research Laboratories, 3-1 Wakamiya Morinosato, Atsugi-shi, Kanagawa, 243-0198 Japan*

[*]Contributed equally to this work

Correspondence should be addressed to gkoster@stanford.edu [G.K.]

**We have grown and studied high quality SrRuO$_3$ films grown by MBE as well as PLD. By changing the oxygen activity during deposition we were able to make SrRuO$_3$ samples that were stoichiometric (low oxygen activity) or with ruthenium vacancies (high oxygen activity). Samples with strontium vacancies were found impossible to produce since the ruthenium would precipitate out as RuO$_2$. The volume of the unit cell of SrRuO$_3$ becomes larger as more ruthenium vacancies are** **introduced. The residual resistivity ratio (RRR) and room temperature resistivity were found to systematically depend on the volume of the unit cell and therefore on the amount of ruthenium vacancies. The RRR varied from ~30 for stoichiometric samples to less than two for samples that were very ruthenium poor. The room temperature resistivity varied from 190 $\mu\Omega$ cm for stoichoimetric samples to over 300 $\mu\Omega$ cm for very ruthenium poor samples. UPS spectra show a shift of weight from the *coherent* peak to the *incoherent* peak around the Fermi level when samples have more ruthenium vacancies. Core level XPS spectra of the ruthenium *3d* lines show a strong screened part in the case of stoichiometric samples. This screened part disappears when ruthenium vacancies are introduced. Both the UPS and the XPS results are consistent with the view that correlation increases as the amount of ruthenium vacancies increase.**

## Introduction

SrRuO$_3$ is a material of uncommon scientific interest. It is unusual among the correlated complex oxides in being a metallic itinerant ferromagnet. It exhibits so-called bad metal behavior at high temperatures but is clearly a Fermi liquid at low temperatures. It is also of widespread utility as a conducting electrode in many devices utilizing complex oxides of various kinds. One of the mysteries of this material is why its transport properties are so sensitive to how it is synthesized, particularly in thin film form. In this paper, using advanced thin film deposition techniques combined with *in situ* photoemission studies, we show that this sensitivity is not a simple matter of defect scattering but, rather, involves a sensitivity of the electronic structure and the associated degree of electronic correlation to the precise stoichiometry [1]. This result is clearly of fundamental importance quite independent of the motivating problem of the dependence of properties of films of SrRuO$_3$ on the means of deposition. Moreover, we discuss how our results can be placed in the context of the degree of correlation in the ruthenate family of complex oxides generally.

The properties observed in the ruthenate family of complex oxides range from very good metals (RuO$_2$) to insulators (Y$_2$Ru$_2$O$_7$). Recently Kim *et al.* [2] suggested that this change in electronic behavior can be attributed to a change in electron-electron correlation. They come to this conclusion based on fits of experimental core level photoemission spectra (X-ray Photoemission Spectroscopy (XPS), Cox et al. [3]) using Dynamic Mean Field Theory [4]. The *3d* peaks of ruthenium are compared for ruthenates put in order of metallicity, and a systematic relative shift in spectral weight from the so called *screened* peak to the *unscreened* peak is observed. To explain these shifts, these authors [2] use a model in which the fitting parameters are the Hubbard *U* and the bandwidth *W* of the ruthenium *4d* band. The outcome of their analysis is that the ratio *U/W* correlates with the ratio of the *screened* and *unscreened* peaks, from which one can deduce that the stronger the *screened* peak the less correlation (i.e., the more metallic).

In UPS spectra, a similar shift in spectral weight is observed. In this case, the $t_{2g}$ peak, which is very close to the Fermi level and referred to as the *coherent* peak (i.e., the quasiparticle band near $E_F$), is reduced as correlation increases, and a broad peak around 1.5 eV starts to rise. This latter peak is called the *incoherent* peak (i.e., the remnant of the Hubbard bands 1-2 eV above and below $E_F$). Both SrRuO$_3$ and CaRuO$_3$ can be classified by their XPS and UPS spectra in this way, despite, for example, similar transport properties at room temperature. Based on the classification by XPS spectra $CaRuO_3$ would be a more correlated system than $SrRuO_3$, however the samples studied in [2] could have suffered from the same off-stoichiometry as we will discuss in this paper (for instance due to the surface or sample preparation processes).

As stated above, here we show that similar systematic variations in the degree of correlation can occur within a single ruthenate as a function of disorder/off-stoichiometry. In particular, it turns out that one source of disorder/off-stoichiometry can be varied in $SrRuO_3$ thin film by changing the deposition conditions or (what turns out to be more or less equivalent) the deposition technique. Specifically, we demonstrate that variation of vacancies on the ruthenium site gives rise to a systematic change in the degree of correlation. Moreover, the transport properties of our samples are clearly linked to their photoemission spectra (XPS and UPS) and to the crystal unit cell parameters. $SrRuO_3$ appears to be a system where these effects of correlation can be studied in a systematic fashion, usually not easily accessible, but we suspect that the underlying physics is generic.

**Experimental**

The thin film samples reported in this paper are grown by two different methods: Molecular Beam Epitaxy (MBE) and Pulsed Laser Deposition (PLD). The samples are grown in the same vacuum chamber with a background pressure of $10^{-9}$ Torr. The MBE samples are deposited in an oxygen background of $10^{-5}$ Torr and at a substrate temperature of 700 °C. All films are grown on $TiO_2$ terminated $SrTiO_3$ substrates, prepared according to Koster *et al.* [5]. Samples are grown at a rate of about 1 Å/s from separate ruthenium and strontium sources, which enables us to vary the ratio of ruthenium to strontium. Electron Impact Emission Spectroscopy (EIES) is used to control the deposition rate of the electron beam heated sources. Typical thickness of the films grown ranges from 200 to 300 Å. During growth, atomic oxygen is provided by an Astex SXRHA microwave plasma source. By adjusting oxygen flow and generator wattage (200-600 W), the amount of atomic oxygen can be controlled [6]. During growth, RHEED is used to monitor the morphology of the samples.

For PLD, a KrF excimer laser produces a 248 nm wavelength beam with typical pulse lengths of 20-30 ns. A rectangular mask shapes the beam, and a variable attenuator permits variation of the pulse energy. The energy density on the target is kept at approximately 2.1 J/cm$^2$. Before each run, the rotating stoichiometric SrRuO$_3$ target is pre-ablated for two minutes at 4 Hertz. Films are deposited with a laser repetition rate of 4 Hertz, with the substrate temperature at 690 °C. The oxygen background pressure was $10^{-5}$ Torr as in the MBE case, however no atomic oxygen was provided for the PLD films. The thickness of the films is the same as for the MBE samples.

X-Ray Diffraction (XRD) experiments were performed on a Philips Xpert thin film diffractometer, to determine both the thickness as well as the crystalline quality of the samples. UPS measurements were performed *in situ* in a vacuum chamber attached to the main growth chamber, which has a base pressure of $<5\times10^{-10}$ Torr, with a VG Scientific ESCAlab Mark II system (non-monochromatic, helium discharge). XPS measurements were performed *in situ* with the same system (non-monochromatic AlKα). We present only *in situ* XPS results, since the surface of SrRuO$_3$ is known to change when exposed to air [7], in a way that gives rise to surface states in the spectra. Electrical transport properties were measured with a Quantum Design Physical Property Measurement System (PPMS).

By varying the deposition conditions, we found that the properties of the films changed. The films so obtained were divided into three groups: ruthenium rich, near stoichiometric and ruthenium poor. The ruthenium rich samples are typically made by using a relatively low oxygen partial pressure and/or low atomic oxygen flux and depositing a little excess ruthenium. The nearly stoichiometric samples are single phase and are obtained by tuning the oxygen activity together with the use of a Ru/Sr ratio close to one. The ruthenium poor samples are created by using relatively high partial oxygen pressures and a Ru/Sr ratio greater than or equal to one (the Ru flux does not seem to be the determining factor here).

It turns out to be impossible to grow $SrRuO_3$ with strontium vacancies. It is possible though to grow samples that are stoichiometric or with ruthenium vacancies (for reasons discussed later). These samples are single phase and have the $SrRuO_3$ crystal structure with no precipitates. The precise stoichiometry is difficult to determine, but, based on the combined analysis (XRD, XPS), we can make the above distinction without further specifying the exact composition of individual samples in each group. All three sample types can be obtained by MBE, but samples made by PLD using a stoichiometric target typically are Ru poor. In the remainder of this paper, we will focus on samples that are either (near) stoichiometric or ruthenium poor.

**Results**

We first present our resistivity measurements. In Fig. 1, the normalized resistance is plotted as a function of temperature for three samples: a stoichiometric (MBE), a ruthenium poor (MBE), and a ruthenium poor (PLD) film. The normalization factor is the room temperature resistivity. For the stoichiometric samples the room temperature resistivity is around 190 µΩ cm, which compares well to values found in literature for polycrystalline and single crystal samples (150 to 200 µΩ cm). When ruthenium vacancies are introduced, the value increases markedly to about 300 µΩ-cm. Also of interest is the Residual Resistivity Ratio (RRR), which is defined as the resistivity at 300 K divided by the value at 4 K. For the best single crystals, values between 50 and 100 are reported [8], and similar results can be obtained for optimized thin films [9]. Like the room temperature resistivity, the RRR is very sensitive to changes in stoichiometry. For the stoichiometric sample in Fig. 1 (black line), the RRR is 26, which is excellent for the thickness of these samples. Whereas for both ruthenium poor films (orange and red lines), the RRR drops to around 7. Note that for the majority of PLD films in the literature, values of 5 or less are reported [10, 11].

Next we turn to the magnetism of our films. The transition temperature, $T_C$, in bulk samples is 160 K, whereas in all of our thin film samples, $T_C$ is reduced by about 10 K [12] due presumably to strain (see the inset in Fig. 1 where the temperature derivatives of the resistance near the transition are compared). From literature, we know that when a sufficient amount of ruthenium vacancies are introduced, $T_C$ is lowered further [1]. This gives us some indication of the order of magnitude of the vacancy density. On this basis we estimate that for the range of samples we studied the vacancy concentration is much smaller than a few percent.

Next we consider the change in lattice parameters on going from the stoichiometric to Ru poor samples. In Fig. 2 a) we show θ/2θ scans for two thick films (310 nm), one grown with MBE in low oxygen pressure and one with PLD in high oxygen pressure, to illustrate the difference in the d-spacing for the out-of-plane (110) direction of $SrRuO_3$ when different deposition techniques are used. In addition, these scans reveal the very high degree of crystallinity of the samples by the existence of finite size fringes in the scans for both samples. As can be seen, the out-of-plane lattice constant is larger for the film grown in high oxygen pressures, explained by a lower ruthenium content [1]. Since the c-axis (in-plane) of the $SrRuO_3$ is fixed by epitaxy (the films are fully strained for all sample thicknesses in this paper), the increase in the (110) direction (linear combination of the a- and b-axis) is directly proportional to an increase in unit cell volume.

We obtained the lattice parameters by refining on six reflections. For samples which were classified as having low ruthenium content, we find a unit cell with larger volume. This increase in the case of ruthenium vacancies is consistent with earlier measurements done on ruthenium deficient samples [1]. The bulk values taken from literature are a=5.572 Å, b=5.533 Å, c=7.849 Å and $V_{bulk}$=242.0 Å$^3$ [1] for single crystals and correspond to an unstrained lattice, unlike the films grown here on $SrTiO_3$.

In Fig. 2 b) we show how changes in the transport data correlate with the volumes of the orthorhombic unit cell for various films studied. Both the room temperature resistivities as well as the RRR show a clear trend with smaller volumes showing the lowest resistivities and highest RRR. These films are typically found in our second group made by MBE (oxygen activity tuned to optimize oxidation and ruthenium sticking). On the other end, PLD films appear to have larger volumes and poorer transport properties. Note that the Curie temperature should also show a similar trend, although the change in $T_C$ is small as shown in Fig. 1.

To get a better understanding of how or whether the electronic properties vary as a function of stoichiometry, we measured the UPS and XPS spectra of some of our samples. In Fig. 3a, the UPS spectra are plotted for the same samples that were used for the transport data in Fig. 1. First, for the stoichiometric $SrRuO_3$ sample grown by MBE, the spectrum shows a peak at the Fermi energy, corresponding to the ruthenium $t_{2g}$ band and a valley at binding energy ~1.5 eV. This peak has been observed before by Kim and coworkers [13] and also is expected based on models [14]. In ruthenium poor samples, as can be seen in Fig. 3a, for both the PLD film, as well as the ruthenium poor MBE film a second, broad peak appears at ~1.5 eV, the so-called *incoherent* peak.

Examples of XPS core level spectra for one stoichiometric and one ruthenium poor sample are plotted in Fig. 3b. For clarity, we reduced these spectra of the ruthenium *3d* doublet by subtracting the strontium $3p_{1/2}$ peak fitted with a Gaussian, which overlaps at lower binding energy. Note that we chose to depict data obtained *in situ* to avoid carbon contamination and its complications near the surface region [7] and which normally also overlaps with Ru *3d*, sacrificing on resolution. The stoichiometric sample shows more spectral weight for the Ru $3d_{3/2}$ peak at lower binding energy compared to the Ru-poor sample. Peak fitting using 3 sets of Gaussians (the area of the peaks in each spin orbit doublet pair are ratio-ed 2:3) shows that at a binding energy of ~278 eV and extra curve (yellow solid) is necessary to fit the data for stoichiometric $SrRuO_3$, which is arguably the so-called *screened* peak typical for $SrRuO_3$. On the other hand, the data for the Ru-poor sample is dominated by the *unscreened* peak.

**Discussion**

Before discussing these data explicitly, we would first like to point out that the stoichiometry of the samples is extremely dependant on the oxygen activity during deposition. For MBE this can be independently varied by controlling the flux of molecular or atomic oxygen. Apparently the sticking of ruthenium (possibly through the formation of $RuO_4$, which is very volatile [15, 16] is a function of oxygen activity: at relatively low oxygen activity the stoichiometry is mostly determined by the supplied Sr/Ru ratio (when excess Ru is supplied, RHEED, SEM and TEM reveal precipitation of $RuO_2$ and the Ru *3d* core-level spectra show very strong *screened* peaks typical for $RuO_2$); at intermediate oxygen activity it is most favorable to achieve perfect stoichiometry (RHEED shows two dimensional growth and XPS/UPS show almost perfect spectra typical for $SrRuO_3$; the 'best' values for resistivity and RRR are obtained here); finally at high oxygen activity, Ru vacancies are unavoidable, independent of the ratio Sr/Ru in the supplied vapor (RHEED shows two dimensional growth, layer-by-layer or even step flow [17], XRD clearly shows a modification in the unit cell indicative of Ru vacancies [1]). We believe the last scenario also to be the case for PLD films, where a high (atomic) oxygen pressure exists within the plume (additional to the background activity) and very little can be done to avoid this. PLD offers the advantage of making the same quality film every time, albeit Ru poor. On the other hand, we have showed that even when ruthenium vacancies are present the crystallinity of the material remains the same compared to the stoichiometric films and therefore cannot be used to explain any of the observed variations in transport and PES. There seems no indication that oxygen vacancies play a significant role but we cannot distinguish this from the formation of ruthenium vacancies by looking the response of lattice constants only. In a study of similar ruthenium deficient samples, oxygen vacancies were not observed [1].

Resistivity is both affected at low temperatures with lower RRR values and at higher temperatures with a variation in lower room temperature resistivity for stoichiometric samples. At low temperatures defect scattering dominates transport properties resulting in the systematic behavior of the RRR values, decreasing almost exponentially with unit cell volume (more defects). On the other hand, at room temperature, resistivity increases more linearly with the unit cell volume. It would be interesting to study the high temperature behavior (beyond the Ioffe-Regel limit [18]) as a function of stoichiometry, which is part of a future endeavor.

Now we turn to the observed differences in PES and their correlation with transport properties as a function of Ru stoichiometry. First, it is logical to exclude the Ru-rich samples from this comparison because both transport and PES would be a mixture of $SrRuO_3$ and precipitated $RuO_2$. Both UPS and XPS spectra of films with vacancies are distinctly different from their stoichiometric counterparts. Our data of the Ru 3$d$ peaks in $SrRuO_3$ with vacancies suggest that these ruthenium poor samples show almost no *screened* peaks. This effect is most associated with the material becoming more correlated, which has been quantified by Kim and coworkers [2]; see also our earlier discussion. In the same spirit, Ru-poor samples show spectral weight at ~1.5 eV in HeI

UPS, which has previously been attributed to an *incoherent* peak in a picture of a strongly-correlated system.

When the ~1.5 eV peak is absent, the $t_{2g}$ peak at the Fermi level becomes more pronounced. It is noteworthy here that the $t_{2g}$ peak at the Fermi level has not yet been observed with scraped bulk samples, indicating that thin-film specimens may provide a better opportunity in investigating an intrinsic electronic structure of $SrRuO_3$ with photoemission spectroscopy as has been pointed out by Kim et al. [13]. The thickness dependence of the electronic structure has been studied by Toyota *et al.* [19] Band structure calculations predict a $t_{2g}$ peak at the Fermi level, which is consistent with the proposition that correlation effects are not as important in this case [20].

**Conclusions**

We have grown $SrRuO_3$ thin films on $SrTiO_3$ under various conditions and conclude that this material exhibits a range of properties due to a subtle change in stoichiometry on the Ru site related to the oxidation conditions during deposition. Resistivity, X-ray diffraction, UPS and XPS all seem to indicate that this change in behavior is due to a changing electron-electron correlation.

**Acknowledgements**

One of us (G.K.) thanks the Netherlands Organization for Scientific Research (NWO, VENI). W.S. thanks the Nanotechnology network in the Netherlands, NanoNed. We would also like to thank M. Naito, Kookrin Char, Ann Marshall, Jim Reiner, Guus Rijnders, Ted Geballe and Robert Hammond for helpful discussions. This work was supported by the DoE BES with additional support from EPRI.

1313

**Figure Captions**

FIG 1. Resistance behavior as a function of temperature for an MBE grown (black solid line), a ruthenium poor MBE grown (red dashed line), and a PLD grown (orange dotted) SrRuO$_3$ film. The resistance has been normalized with respect to the values at 300 K. The inset shows the derivative of each line, $T_c$ has been set as the point where the derivative is maximal. The resistivity ($\rho$) of the stoichiometric MBE grown SrRuO$_3$ at room temperature is about 190 μΩ cm and its carrier density at 4 K is $2 \times 10^{22}$ cm$^{-3}$.

FIG 2. a) Comparison of θ/2θ scans of two thick (310 nm) films grown by PLD and MBE. The fringes indicate good crystalline quality of the films, the shift in peak position indicates that the PLD film has a larger out of plane lattice parameter (110 direction). b) A comparison between the volume of the SrRuO$_3$ unit cell in thin films on SrTiO$_3$ substrates grown by either MBE or PLD and their transport properties: room temperature resistivity on the right y-axis (blue) and residual resistance ratio on the left y-axis (black). The dashed lines are guides to the eye.

FIG 3. a) UPS spectra of three SrRuO$_3$ films: an MBE film, a ruthenium poor MBE film and a PLD grown film. All samples show states at the Fermi edge and a Ru t$_{2g}$ peak close to the Fermi level. b) XPS detail spectra of the Ru 3d doublet of a stoichiometric film (top) and a ruthenium poor film (bottom), including fitting of the spectra using 3 spin orbit coupled doublets represented by Gaussians; the overlapping Sr2p$_{1/2}$ peak (Gaussian) was subtracted for both spectra. In the spectrum of stoichiometric film more pronounced

screened peaks (yellow) are needed to give a good fit, indicative of less correlated behavior. For the ruthenium poor spectrum the screened peaks are almost non-existent.

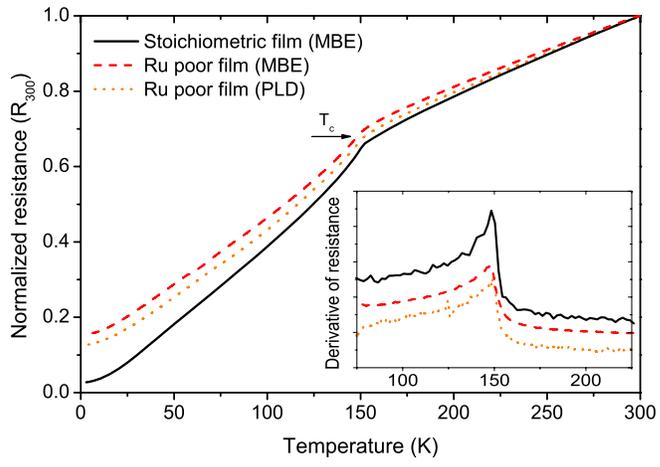

Siemons et al. Figure 1.

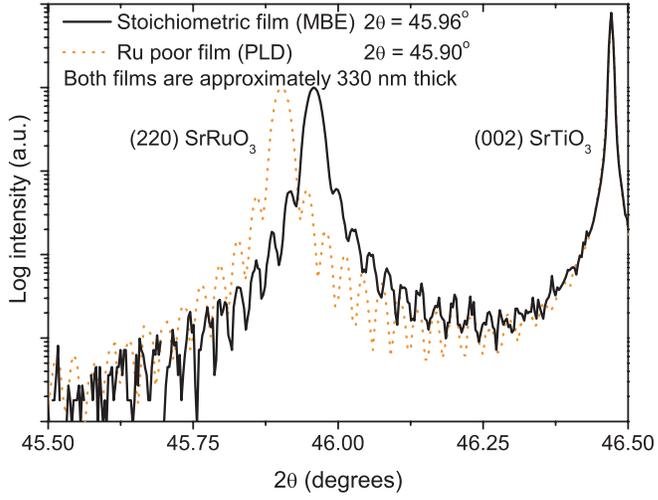

Siemons et al. Figure 2 a)

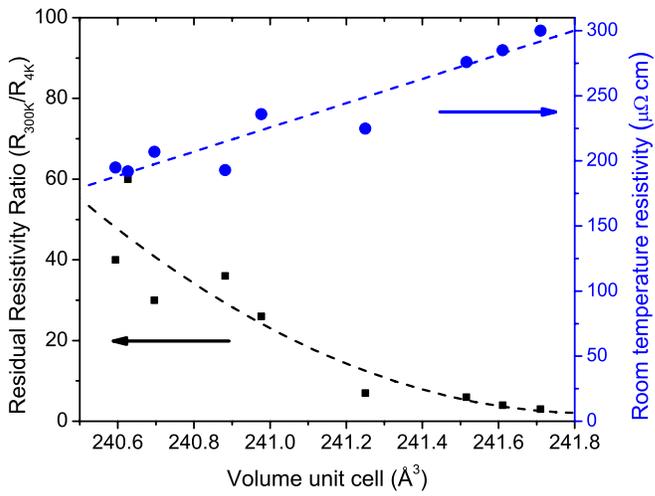

Siemons et al. Figure 2 b).

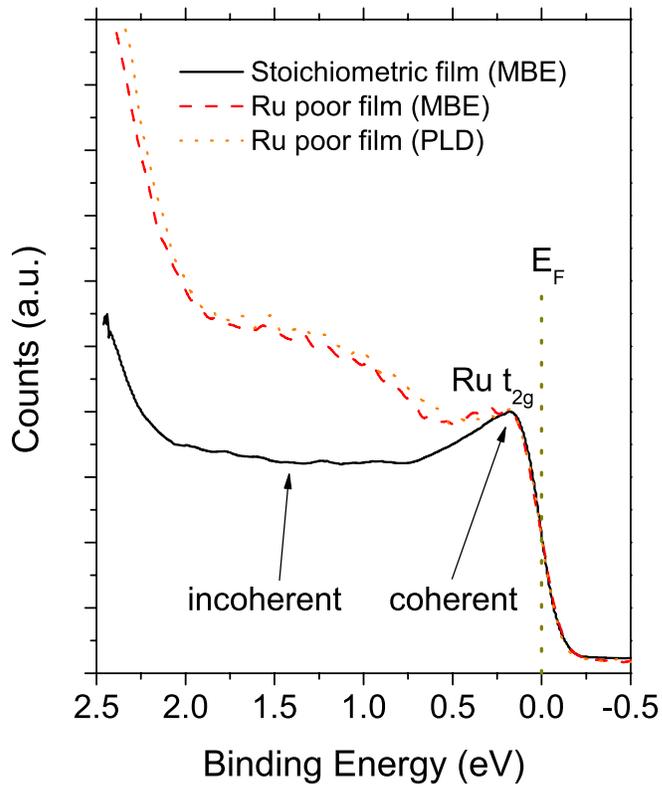

Siemons et al. Figure 3 a)

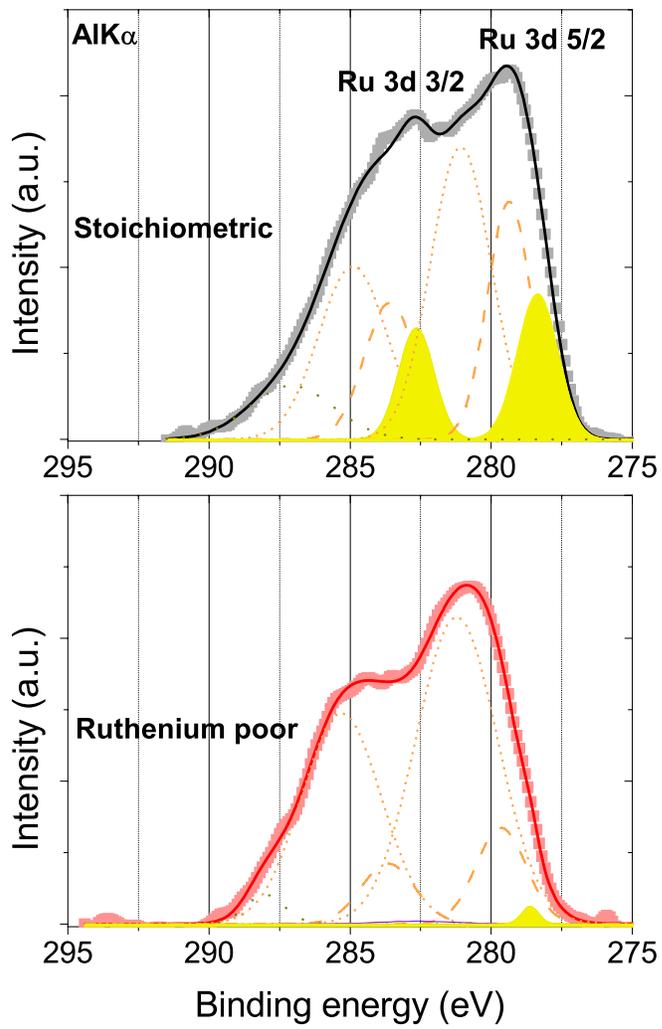

Siemons et al. Figure 3 b)